 \documentclass[final,authoryear,5p,times,twocolumn]{elsarticle}

\usepackage{booktabs}
\usepackage{amssymb}
\usepackage{amsmath}
\usepackage{cases}

\usepackage[nonumberlist, toc, acronym]{glossaries}
\usepackage{acronym} 

\journal{CES} 

\begin{document}

\begin{frontmatter}



\title{Lattice Boltzmann based discrete simulation for gas-solid fluidization}


\author[label1]{Limin Wang \corref{cor1}}
\ead{lmwang@home.ipe.ac.cn} \cortext[cor1]{Corresponding author.
Tel.: +86 10 8254 4942; fax: +86 10 6255 8065.}
\author[label1,label2]{Bo Zhang}
\author[label1]{Xiaowei Wang}
\author[label1]{Wei Ge}
\author[label1]{Jinghai Li}

\address[label1]{The EMMS Group, State Key Laboratory of Multiphase Complex Systems, Institute of Process Engineering, Chinese Academy of Sciences, Beijing 100190, China}
\address[label2]{State Key Laboratory of Organic-Inorganic Composites, Beijing University of Chemical Technology, Beijing 100029, China}

\begin{abstract}
Discrete particle simulation, a combined approach of computational fluid dynamics and discrete methods such as DEM (Discrete Element Method), DSMC (Direct Simulation Monte Carlo), SPH (Smoothed Particle Hydrodynamics), PIC (Particle-In-Cell), etc., is becoming a practical tool for exploring lab-scale gas-solid systems owing to the fast development of parallel computation. However, gas-solid coupling and the corresponding fluid flow solver remain immature. In this work, we propose a modified lattice Boltzmann approach to consider the effect of both the local solid volume fraction and the local relative velocity between particles and fluid, which is different from the traditional volume-averaged Navier-Stokes equations. A time-driven hard sphere algorithm is combined to simulate the motion of individual particles, in which particles interact with each other via hard-sphere collisions, the collision detection and motion of particles are performed at constant time intervals. The EMMS (energy minimization multi-scale) drag is coupled with the lattice Boltzmann based discrete particle simulation to improve the accuracy. Two typical fluidization processes, namely, a single bubble injection at incipient fluidization and particle clustering in a fast fluidized bed riser, are simulated with this approach, with the results showing a good agreement with published correlations and experimental data. The capability of the approach to capture more detailed and intrinsic characteristics of particle-fluid systems is demonstrated. The method can also be used straightforward with other solid phase solvers.
\end{abstract}


\end{frontmatter}

\section{Introduction}
\label{}
Gas-solid fluidization systems are widely encountered in both physical and chemical processes for many industries, for instance, fluid catalytic cracking (FCC), circulating fluidized bed combustion (CFBC), coal gasification, and sulfide roasting. Earlier studies of these systems mainly focused on experimental investigations including measurement of macroscopic hydrodynamic behavior and development of some corresponding correlations. In recent decades, to quantitatively understand the complex hydrodynamics of gas-solid fluidization, the computational fluid dynamics approach is adopted in many cases, and a lot of numerical methods in the hydrodynamic modeling and simulation of gas-solid fluidization have been proposed, such as two-fluid model (TFM) (Anderson and Jackson, 1967; Ishii, 1975), quadrature-based moment methods (QBMM) (Fox, 2008, 2009a,b; Desjardins et al., 2008;), discrete particle simulation (DPS) (Tsuji et al., 1993; Hoomans et al., 1996; Xu and Yu, 1997), and direct numerical simulation (DNS) (Hu et al., 1992; Ma et al., 2006; Wang et al., 2010; Xiong et al., 2012).

Among these numerical methods, the most frequently used TFM treats the gas and solid phases as two interpenetrating continua, and locally averaged quantities such as volume fractions, velocities, species concentrations, and temperatures of gas and solid phases appear as dependent field variables (Anderson and Jackson, 1967; Ishii, 1975). To derive TFM using ensemble averaging techniques, terms such as effective stresses and the inter-phase interaction have to be introduced, which require constitutive equations for closure. Only under very limited conditions, those constitutive equations can be obtained rigorously from the kinetic theory of granular flow (Gidaspow, 1994), otherwise we have to resort to empirical models. The accuracy and effectiveness of TFM are, therefore, still unsatisfactory in many circumstances. The recently developed QBMM permits to solve population balance equation (PBE) in commercial CFD codes at relatively low computational cost. However, its application to the context of multiphase flows is to be explored (Mazzei, 2011). Comparably, DNS not only fully resolves the motion of each individual solid particle and fluid flow, but also directly calculates the hydrodynamic force acting on each individual solid particle from the stress on the fluid-solid boundary. Due to its capability in detailed solution around each particle, DNS has been regarded as the most accurate method for the simulation of gas-solid flow. Unfortunately, DNS is too costly for predicting the hydrodynamics in large industrial scale fluidized beds even at low Reynolds numbers, let alone the high Reynolds number cases where their grid size and time step are limited by the Kolmogorov length scale and the turbulence time scale (Xiong et al., 2012).

For numerical modeling of gas-solid fluidized beds mentioned above, TFM is computationally more economic but inaccurate, while DNS is computationally more accurate but expensive. So it is natural to ask whether there exists a better alternative combining the advantages of the two methods for modeling the gas-solid flows. As a particle-scale approach, DPS is somehow in between these two ends and seems to give a good balance among accuracy, cost and efficiency. Specifically, DPS resolves the continuum fluid flow at the scale of computational cells in CFD, describes the motion of individual particles by the well-established Newton's equations of motion, and models particle-particle interactions through different collision models such as the hard-sphere model and the soft-sphere model, which has been proven to be effective in modeling various particle flow systems (Deen et al., 2007; Zhu et al., 2007, 2008), such as slugging fluidized bed (Xu et al., 2007), spouted bed (Zhao et al., 2008), pneumatic conveying (Kuang et al., 2008), bubbling fluidized bed (Geng and Che, 2011), sound-assisted fluidized bed (Wang et al., 2011), and cyclone separator (Chu et al., 2011). However, in all these mentioned work, the fluid motion with suspended solids is commonly governed by the volume-averaged Navier-Stokes equations or their simplified forms (Tsuji et al., 1993; Hoomans et al., 1996; Xu and Yu, 1997; Mikami et al., 1998), and those equations are solved based on implicit schemes no matter by Fluent (Chu and Yu, 2008a, b; Chu et al., 2009a, b, 2011; Wu et al., 2010; Zhao et al., 2010), OpenFOAM (Su et al., 2011; Goniva et al., 2012), MFIX (Darabi et al., 2011; Garg et al., 2012; Li and Guenther, 2012; Li et al., 2012a,b; Gopalakrishnan and Tafti, 2013), or in-house codes (Ouyang and Li, 1999a, b; Zhou et al., 2004a, b; Zhao et al., 2009; Wang et al., 2009; Wu et al., 2009). With implicit methods the discretized equations are solved simultaneously which inevitably requires some kind of global data dependence and hence global communication. Therefore, most algorithms involved suffer from relatively lower scalability and parallel efficiency, which becomes a grand challenge for fast simulation of large-scale industrial systems.

As a smoothed alternative to lattice gas automata (LGA), lattice Boltzmann method (LBM) (McNamara and Zanetti, 1988) is an efficient second-order flow solver capable of solving various systems for hydrodynamics owing to its explicit solution of particle distribution function, algorithmic simplicity, natural parallelism, and flexibility in boundary treatment (Chen and Doolen, 2003). Therefore, LBM becomes an increasingly popular approach to simulation of complex flows (Aidun and Clausen, 2010) and can be easily incorporated into DPS. Filippova and Hanel (1997) proposed a combination of lattice-BGK model and Lagrangian approach, and performed three-dimensional simulation of gas-particle flow through filters with one-way coupling, where the fluid affected the particles but the particles did not affect the fluid. Chen et al. (2004) simulated particle-laden flow over a backward-facing step with two-way coupling, where a modified lattice-BGK model was developed for the fluid flow and a Lagrangian approach for particles. But they did not consider the effect of solid volume fraction on gas flows. Sungkorn et al. (2011) proposed a gas-liquid Lagrangian-LBM to simulate turbulent gas-liquid bubbly flows with a relatively low gas holdup. Specifically, they solved the continuous liquid phase by single-phase lattice Boltzmann equation (LBE) incorporated with large eddy simulation (LES) (Smagorinsky, 1963), and evolved the dispersed gas phase (i.e. the individual bubbles) by Lagrangian trajectories, but did not include the gas volume fraction in the conservation equations and its effect on drag force.

In this paper, we proposed a modified LBE to model the fluid flow and developed the corresponding fluid-solid interaction model in the framework of DPS. The effects of both the local solid volume fraction and the local relative velocity between particles and fluid are considered. The equations of motion governing individual particles are solved with time-driven hard-sphere (TDHS) model. It is noteworthy that the computational strategy herein has ever been implemented in direct simulation of particle-fluid systems (Wang et al., 2010) where the modified LBE was used with particle size much larger than the cell spacing. In the present work, the partial saturation concept has been extended to model the objects much smaller than the cell spacing (i.e. porous media), and both the linear and nonlinear drag effects of the solid phase (media) have been considered in the lattice Boltzmann based discrete particle simulation for the first time.

\section{Numerical approach}
\label{}
The objective of this research is to develop a lattice Boltzmann based numerical method for discrete simulation of gas-solid fluidization systems. For illustration, we used two-dimensional nine-velocity (D2Q9) lattice Boltzmann model as an example, and the solid particles distributed in the lattice cell are described by the time-driven hard sphere model. A schematic diagram of this method is shown as Fig. 1.

\begin{figure*}[htb]
  \centering
  \includegraphics[width=14.3cm]{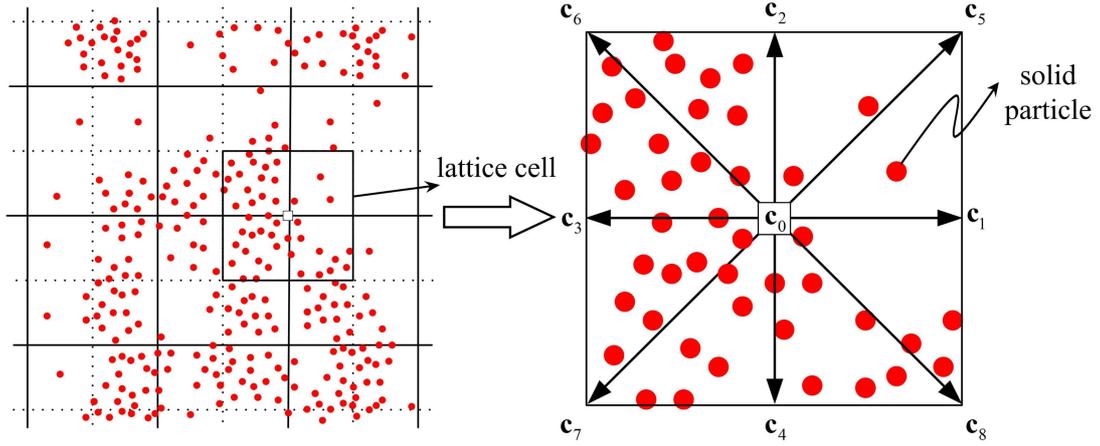}
  \caption{The schematic diagram of D2Q9 computational fluid lattices and solid particles.}\label{fig1}
\end{figure*}

\subsection{LBM for gas phase hydrodynamics}
The particle distribution function $f({t,{\bf{x}},{\bf{v}}})$ is the number of particles per unit volume with velocity $\bf{v}$ at time t and position $\bf{x}$, which is often used in non-equilibrium statistical mechanics. The evolution of the distribution function f is described by the Boltzmann equation (Chapman and Cowling, 1970):
\begin{equation}
\frac{{\partial f({t,{\bf{x}},{\bf{v}}})}}{{\partial t}} + {\bf{v}}\frac{{\partial f( {t,{\bf{x}},{\bf{v}}})}}{{\partial {\bf{x}}}} = \Omega
\end{equation}
The left-hand side of this equation is an advection term, while the right-hand side contains the collision operator $\Omega$ which describes the interaction between particles. External forces are neglected in this derivation. After a finite difference discretization of the Boltzmann equation in time, space and velocity, LBE is simplified as (McNamara and Zanetti, 1988),
\begin{equation}
  {f_i}({\bf{x}} + {{\bf{c}}_i}\Delta t,t + \Delta t) - {f_i}({\bf{x}},t) = \Omega _i
\end{equation}
where the subscript \textit{i} represents the orthogonal and diagonal directions in the Cartesian coordinates, $f_i$ is the particle distribution function in direction \textit{i}, $\mathbf{c}_{i}$ is the discrete velocity in direction \textit{i}, \textbf{x} and \textit{t} are the lattice simulation position and time, respectively, and $\Omega_i$ is collision term in direction \textit{i}. Assuming that the collision frequency is a function of spatial coordinates and time, and independent of the molecular velocity, the collision term $\Omega_i$ can be approximated by the so-called BGK (Bhatnagar et al., 1954) model:
\begin{equation}
  \Omega _i^{} =  - \frac{{\Delta t}}{\tau }\left( {{f_i} - f_i^{{\rm{eq}}}} \right)
\end{equation}
Here $\tau  = 3\nu  + \frac{{\Delta t}}{2}$ is the relaxation time related to kinematic viscosity and the discrete time step $\Delta t$, $f_i^{{\rm{eq}}}$ is the equilibrium distribution function expressed as (Qian et al., 1992)
 \begin{equation}
  f_i^{\rm{eq}}(\rho,{\bf{u}})=\omega_{\it{i}} \rho \left( 1 + \frac{{\bf{c}}_{\it{i}} \cdot {\bf{u}}}{c_{\rm{s}}^{\rm{2}}} + \frac{({\bf{c}}_{\it{i}} \cdot {\bf{u}})^2}{2c_{\rm{s}}^{\rm{4}}}-\frac{{\bf{u}} \cdot {\bf{u}}}{2 c_{\rm{s}}^{\rm{2}}}\right)
 \end{equation}
where $\mathbf{c}_i$ is defined as ${{\bf{c}}_i}{\rm{ = }}\left\{ {\left. {\cos \left[ {\left( {i - {\rm{1}}} \right)\pi {\rm{/}}2} \right],{\kern 3pt} \sin \left[ {\left( {i - {\rm{1}}} \right)\pi {\rm{/}}2} \right]} \right\}} \right.$ for $i{\rm{=1\sim4}}$ and ${{\bf{c}}_i}{\rm{ = }}\sqrt 2 \left\{ {\left. {\cos \left[ {\left( {i - 5} \right)\pi {\rm{/}}2} \right],{\kern 3pt} \sin \left[ {\left( {i - 5} \right)\pi {\rm{/}}2} \right]} \right\}} \right.$  for $i{\rm{=5\sim8}}$, and ${{\bf{c}}_0}{\rm{ = }}0$. The weights are given by ${\omega_0}{\rm{ = 4/}9}$, ${\omega_i}{\rm{ = }}1{\rm{/}}9$ for $i{\rm{=1\sim4}}$, ${\omega_i}{\rm{ = }}1{\rm{/}}36$ for $i{\rm{=5\sim8}}$, and $c_{\rm{s}}$ is the speed of sound and equals to $\sqrt{3}\rm{/}3$  in lattice unit, $\rho$ and \textbf{u} are density and velocity, respectively.

The macroscopic quantities such as mass density and momentum density can then be obtained by evaluating the hydrodynamic moments of the distribution function $f_{i}({\bf{x},t})$, namely,
\begin{equation}
  \rho  = \sum\limits_{i = {\rm{0}}}^8 {{f_i}} {\kern 1pt} {\kern 1pt} {\kern 1pt} {\kern 1pt} {\kern 1pt} {\kern 1pt} {\kern 1pt} {\kern 1pt} {\kern 1pt} {\kern 1pt} {\kern 1pt} {\kern 1pt} {\kern 1pt} {\kern 1pt} {\kern 1pt} {\kern 1pt} {\kern 1pt} {\kern 1pt} {\kern 1pt} {\kern 1pt} {\kern 1pt} {\kern 1pt} {\kern 1pt} {\kern 1pt} {\kern 1pt} {\kern 1pt} {\kern 1pt} {\kern 1pt} {\rm{and}}{\kern 1pt} {\kern 1pt} {\kern 1pt} {\kern 1pt} {\kern 1pt} {\kern 1pt} {\kern 1pt} {\kern 1pt} {\kern 1pt} {\kern 1pt} {\kern 1pt} {\kern 1pt} {\kern 1pt} {\kern 1pt} {\kern 1pt} {\kern 1pt} {\kern 1pt} {\kern 1pt} {\kern 1pt} {\kern 1pt} {\kern 1pt} {\kern 1pt} {\kern 1pt} {\kern 1pt} {\kern 1pt} {\kern 1pt} {\kern 1pt} {\kern 1pt} {\kern 1pt} {\kern 1pt} {\kern 1pt} {\kern 1pt} \rho {\bf{u}} = \sum\limits_{i = {\rm{0}}}^8 {{{\bf{c}}_i}{f_i}}
\end{equation}

By the Chapman-Enskog procedure (Sterling and Chen, 1996; Guo et al., 2002), the incompressible Navier-Stokes equations can be obtained from LBE in the limit of small Mach number. Therefore, the standard LBE is just suitable for single-phase flow. To model incompressible gas flow through suspended particles, an immersed moving boundary method for a medium with porosity needs to be introduced in LBM framework.

The key idea of the immersed moving boundary method is that the effect of boundary is modeled using a source term in the momentum equations. Here, the particle-fluid coupling is implemented by immersed moving boundary method (Noble and Torczynski, 1998; Cook et al., 2004), which introduces a term that depends on the percentage of the cell saturated with fluid to modify the collision operator and represent the effect of both the solid volume fraction and the gas-solid slip velocity on hydrodynamics. Although most of applications of the immersed moving boundary are associated with the objects whose size is much larger than the cell spacing (Noble and Torczynski, 1998; Cook et al., 2004; Feng et al., 2007; Wang et al., 2010; Zhou et al., 2011; Xiong et al., 2012; Wei et al., 2013), the method is theoretically not confined to this scale. Therefore, the partial saturation concept is extended to model the objects with much smaller size than the cell spacing, and the corresponding LBE modified for partially saturated cells (Noble and Torczynski, 1998) reads
\begin{equation}
\begin{split}
  {f_i}({\bf{x}} + {{\bf{c}}_i}\Delta t,& t + \Delta t) = {f_i}({\bf{x}},t)  \\
  & - \frac{1}{\tau }\Big(1 - \gamma (\varepsilon _{\rm{s}} ,\tau)\Big)\Big({f_i}({\bf{x}},t) - {f_i}^{{\rm{eq}}}({\bf{x}},t)\Big)\\
  & + \gamma (\varepsilon _{\rm{s}} ,\tau )\Omega _i^{\rm{s}}
\end{split}
\end{equation}
where $\Omega _i^{\rm{s}}$ is an additional collision term that accounts for the bounce back of the non-equilibrium part of the distribution function, and $\Omega _i^{\rm{s}}$ is given by (Noble and Torczynski, 1998)
\begin{equation}
  \Omega _i^{\rm{s}} = {f_{ - i}}({\bf{x}},t) - {f_i}({\bf{x}},t) + f_i^{{\rm{eq}}}(\rho ,{{\bf{v}}_{\rm{s}}}) - f_{ - i}^{{\rm{eq}}}(\rho ,{\bf{u}})
\end{equation}
where $-i$ denotes the component of the distribution function opposite to $i$ and $\mathbf{v}_{\rm{s}}$ is the average velocity of solid phase at the computational lattice. $\mathbf{v}_{\rm{s}}$ can be approximately calculated as
\begin{equation}
  {{\bf{v}}_{\rm{s}}} = \frac{{\sum\nolimits_{k = 1}^{{n_{{\rm{tot}}}}} {{{\bf{v}}_k}} }}{{{n_{tot}}}}
\end{equation}
where $\mathbf{v}_k$ is the velocity of solid particle \textit{k} and $n_{\rm{tot}}$ is the number of solid particles in fluid cell (see Fig.1).

The weighting function $\gamma$ in Eq. (6) depends on the relaxation time $\tau$ and the solid volume fraction $\varepsilon_{\rm{s}}$ of each cell, which is given, among other forms, by (Noble and Torczynski, 1998)
\begin{equation}
  \gamma ({\varepsilon _{\rm{s}}},\tau ) = \frac{{{\varepsilon _{\rm{s}}}(\tau  - 0.5)}}{{(1 - {\varepsilon _{\rm{s}}}) + (\tau  - 0.5)}}
\end{equation}

In our case, since the gas velocity $\bf{u}$ presents a spatial average over areas fully, partially or not occupied by the solid phase, a different expression of $\gamma$ may be developed. But as a rough estimation, we will at first neglect the actual difference of gas velocity at different locations within each lattice and continue to use the expression of Eq. (9). Obviously, $\gamma  = 0$ and $\gamma  = 1$ represent pure fluid and pure solid states, respectively.

The solid volume fraction $\varepsilon_{\rm{s}}$ is related to the porosity of lattice cell  $\varepsilon_{\rm{2d}}$, namely ${\varepsilon _{{\rm{2d}}}} = 1 - {\varepsilon _{\rm{s}}}$. $\varepsilon_{\rm{2d}}$, an important parameter that influences both the gas-phase and solid-phase motions, can be calculated based on the area occupied by the particles in the lattice cell. The two-dimensional porosity of solid phase is given by
\begin{equation}
  {\varepsilon _{{\rm{2d}}}} = 1 - \frac{{{n_{{\rm{tot}}}} \cdot \pi {r^2}}}{{{h^2}}}
\end{equation}
where \textit{r} is the radius of solid particle and \textit{h} is lattice spacing between nodes. As the drag-force correlation is based on 3D systems, we also use a 3D porosity transformed from 2D porosity for this purpose. According to (Ouyang and Li, 1999a),
\begin{equation}
  {\varepsilon _{{\rm{3d}}}} = 1 - \frac{{\sqrt 2 }}{{\sqrt {\pi \sqrt 3 } }}{\left( {1 - {\varepsilon _{{\rm{2d}}}}} \right)^{{3 \mathord{\left/
 {\vphantom {3 2}} \right.
 \kern-\nulldelimiterspace} 2}}}
\end{equation}

To capture turbulent structures in gas-solid fluidization systems which are usually associated with large Reynolds numbers or become turbulent in nature, meso-scale turbulence models are incorporated into the volume-averaged Navier-Stokes equations in DPS (Zhou et al., 2004a, b; Gui et al. 2008; Liu and Lu, 2009), here LES incorporated into LBM is used (Krafczyk et al., 2003; Yu et al., 2005). The effect of the small sub-grid eddies on the large-scale flow structures are modeled through an additional turbulent viscosity ${\nu _{\rm{e}}}$.

In Smagorinsky-based subgrid scale (SGS) model (Smagorinsky, 1963), ${\nu _{\rm{e}}}$ depends on the strain rate:
\begin{equation}
  {\nu _{\rm{e}}} = {\left( {{C_{\rm{s}}}\Delta x} \right)^2}\left\| S \right\|,{\kern 1pt} {\kern 1pt} {\kern 1pt} {\kern 1pt} {\kern 1pt} {\kern 1pt} {\kern 1pt} {\kern 1pt} {\kern 1pt} {\kern 1pt} {\kern 1pt} {\kern 1pt} {\kern 1pt} {\kern 1pt} {\kern 1pt} {\kern 1pt} {\kern 1pt} {\kern 1pt} {\kern 1pt} {\kern 1pt} {\kern 1pt} {\kern 1pt} {\kern 1pt} {\kern 1pt} {\kern 1pt} {\kern 1pt} {\kern 1pt} {\kern 1pt} {\kern 1pt} {\kern 1pt} S = \sqrt {2{S_{ij}}{S_{ij}}}
\end{equation}
Here  $C_{\rm{s}}$ is the Smagorinsky constant and strain rate tensor ${S_{ij}} = {{\left( {{\partial _j}{u_i} + {\partial _i}{u_j}} \right)} \mathord{\left/
 {\vphantom {{\left( {{\partial _j}{u_i} + {\partial _i}{u_j}} \right)} 2}} \right.
 \kern-\nulldelimiterspace} 2}$ can be obtained directly by computing the momentum fluxes $Q_{ij}$, which are second-order moments of the non-equilibrium distribution function (Yu et al., 2005):
 \begin{equation}
 \begin{split}
   &{S_{ij}} = \frac{1}{{2\rho c_{\rm{s}}^{\rm{2}}\tau }}{Q_{ij}},{\kern 12pt} {Q_{ij}} = \sum\limits_{k = 0}^8 {{c_{ki}}{c_{kj}}\left( {{f_k} - f_k^{\rm{eq}}} \right)} \\
   &S = \frac{Q}{{2\rho c_{\rm{s}}^{\rm{2}}{\tau _{\rm{t}}}}},{\kern 25pt} Q = \sqrt {2{Q_{ij}}{Q_{ij}}}
 \end{split}
 \end{equation}
where $c_{ik}$ is the \textit{k}th component of the lattice velocity $\mathbf{c}_i$. Based on an eddy relaxation time assumption for sub-grid scale stress, the effect of the flow structure at the unresolved scale is modeled through an effective collision relaxation time $\tau_{\rm{e}}$, which is the eddy relaxation time with respect to the eddy viscosity $\nu_{\rm{e}}$. The eddy viscosity is then incorporated into the LBE by using $\tau_{\rm{t}}=\tau + \tau_{\rm{e}}$  instead of $\tau$ in Eq. (6). Accordingly,
\begin{equation}
  {\tau _{\rm{t}}} = \frac{3}{{{c^2}}}{\nu _{\rm{t}}} + \frac{1}{2}\Delta t = \frac{3}{{{c^2}}}\left( {\nu  + {\nu _{\rm{e}}}} \right) + \frac{1}{2}\Delta t
\end{equation}
where \textit{c} is the lattice speed given by ${h \mathord{\left/
 {\vphantom {h {\Delta t}}} \right.
 \kern-\nulldelimiterspace} {\Delta t}}$ . From Eq. (13) and Eq. (14), a quadratic equation is obtained, which yields
\begin{equation}
  {\tau _{\rm{t}}} = \tau  + {\tau _{\rm{e}}} = \frac{1}{2}\left( {\tau  + \sqrt {{\tau ^2} + 18C_{\rm{s}}^2{{\left( {\Delta x} \right)}^2}{\kern 1pt} Q} } \right)
\end{equation}

It should be pointed out that Eq. (15) is suitable for both laminar and turbulent flows. On the one hand, under high gas velocity, \textit{Q} is large and hence $\tau_{\rm{t}} \gg\tau$, which expresses the effect of motion at unresolved scale. On the other hand, under low gas velocity, \textit{Q} vanishes and $\tau_{\rm{t}} \approx \tau$, describing the laminar flow.

\subsection{TDHS for particle dynamics}

Each individual solid particle, treated as a point-volume particle, has four properties:  mass (\textit{m}), radius (\textit{r}), position (\textbf{P}), and velocity (\textbf{v}). According to Newton's second law of motion, the equation of motion for solid particle \textit{k} is
\begin{equation}
  m\frac{{d{{\bf{v}}_k}}}{{dt}} = m{\bf{g}} + {\left( {{{\bf{F}}_{\rm{d}}}} \right)_k} + {\left( {{{\bf{F}}_{\rm{c}}}} \right)_k} + {\left( {V}_{\rm{p}} \right)_k} \nabla p_{k}
\end{equation}
where \textbf{g}  is the gravitational acceleration, ${\left( {{{\bf{F}}_{\rm{d}}}} \right)_k}$  is the drag force,  ${\left( {{{\bf{F}}_{\rm{c}}}} \right)_k}$ is the collision force, $\left( {V}_{\rm{p}} \right)_k$ is the particle's volume, and $\nabla p_{k}$ is the pressure gradient.

\begin{figure}[htb]
  \centering
  \includegraphics[width=8.1cm]{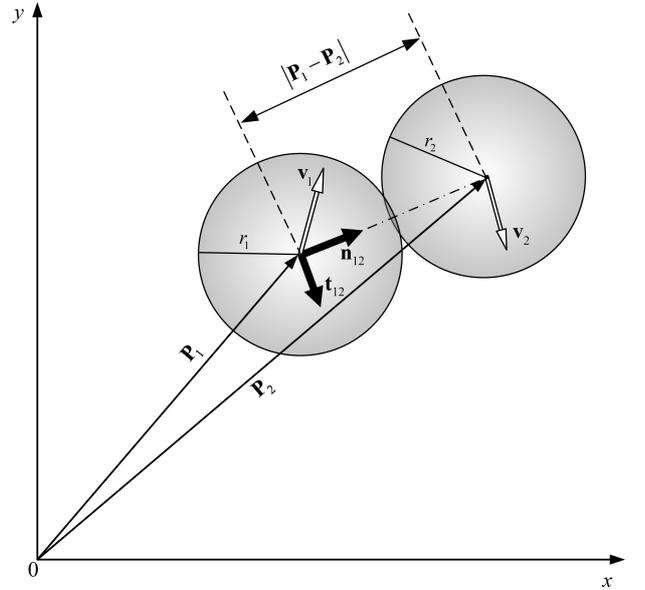}
  \caption{The schematic illustration of two colliding particles.}\label{fig2}
\end{figure}

Time-driven hard-sphere model (Hopkins and Louge, 1991) is used for particle-particle collisions; while particles interact as binary and quasi-instantaneous hard-spheres, and both the collisions are detected and the particle state are updated at constant time intervals. Namely, when two particles are in contact with each other and if the internal product of ${{\bf{P}}_1} - {{\bf{P}}_2}$ and ${{\bf{v}}_1} - {{\bf{v}}_2}$ is negative, which means that the particles are moving closer to each other, they will collide (see Fig. 2). For the collision between particle 1 and particle 2, the post-collisions velocities can be obtained by the following formulae:
\begin{equation}
  {\bf{v}}_1^* = {{\bf{v}}_1} - \frac{{(1 + e){m_2}}}{{{m_1} + {m_2}}}\frac{{({{\bf{v}}_1} - {{\bf{v}}_2}) \cdot ({{\bf{P}}_1} - {{\bf{P}}_2})}}{{{{\left| {{{\bf{P}}_1} - {{\bf{P}}_2}} \right|}^2}}}({{\bf{P}}_1} - {{\bf{P}}_2})
\end{equation}
\begin{equation}
  {\bf{v}}_2^* = {{\bf{v}}_2} + \frac{{(1 + e){m_1}}}{{{m_1} + {m_2}}}\frac{{({{\bf{v}}_1} - {{\bf{v}}_2}) \cdot ({{\bf{P}}_1} - {{\bf{P}}_2})}}{{{{\left| {{{\bf{P}}_1} - {{\bf{P}}_2}} \right|}^2}}}({{\bf{P}}_1} - {{\bf{P}}_2})
\end{equation}
where ``*'' means post-collision velocities, \textit{e} is the restitution coefficient that represents the ratio of speeds after (post-) and before (pre-) collision, and the velocities in the right-hand side of the equations are pre-collision velocities. The walls involved in simulations are composed of infinitely heavy particles with zero velocity, and the collisions between particles and the wall also satisfy Eqs. (17) and (18). In the next time step, the particles move to new positions with their new velocities.

It should also be noted that particle-particle interactions are important in simulating gas-solid fluidization, especially for dense gas-fluidized beds. In time-driven hard-sphere model mentioned above, only the binary-particle collision mechanism and the normal component of contact force are considered, which limits the model to low solid holdup conditions (less than 30$\sim$40\% by volume). For dense gas-fluidized beds, discrete element method (DEM) (Cundall and Strack, 1979) might be expected to reflect both the multi-particle collision mechanism and the tangential component of contact force.

\subsection{Inter-phase momentum transfer}
The drag correlations are of great importance in numerical models that predict the flow behavior of gas-solid fluidized beds. The inter-phase momentum transfer coefficient  based on the traditional Ergun and Wen \& Yu correlations (Gidaspow, 1994; Gidaspow and Jiradilok., 2009) is only valid for homogeneous particle suspensions, and insufficient to capture the heterogeneous structures of gas-solid fluidization. A structure-dependent drag based on the energy minimization multi-scale model (EMMS) is used here, that is (Yang et al., 2003)
\begin{equation}
\beta =\left\{
\begin{aligned}
&\frac{3}{4}\frac{(1-\varepsilon_{\rm{g}}) \varepsilon_{\rm{g}} \rho_{\rm{g}} \left|\bf{u}-\bf{v}\right|}{d_{p}} {C_{\rm{d0}}} \cdot \omega {\rm{,}} &\varepsilon_{\rm{g}}{\rm{ > }}0.74,\\
&150\frac{(1 - {\varepsilon _{\rm{g}}})^2 \mu_{\rm{g}}}{\varepsilon_{\rm{g}}d_{\rm{p}}^2} + 1.75\frac{(1-{\varepsilon_{\rm{g}}})\rho_{\rm{g}}\left|\bf{u} - \bf{v}\right|}{d_{\rm{p}}}{\rm{,}} & \varepsilon_{\rm{g}} \le 0.74.
\end{aligned}\right.
\end{equation}

According to Yang et al. (2003), the drag correction factor
\begin{equation}
\omega =\left\{
\begin{aligned}
&\frac{{0.0214}}{{4{{\left( {{\varepsilon _{\rm{g}}} - 0.7463} \right)}^2}{\rm{ + }}0.0044}}- 0.5760{\rm{,}} &0.74{\rm{ < }}{\varepsilon _{\rm{g}}} \le 0.82,\\
&\frac{{0.0038}}{{4{{\left( {{\varepsilon _{\rm{g}}} - 0.{\rm{7}}789} \right)}^2}{\rm{+}}0.0040}}-0.0101{\rm{,}}&0.{\rm{82 < }}{\varepsilon _{\rm{g}}} \le 0.97,\\
&32.8295{\varepsilon _{\rm{g}}}- 31.8295{\rm{,}} &{\varepsilon _{\rm{g}}}{\rm{ > }}0.97.
\end{aligned}\right.
\end{equation}

The drag coefficient $C_{\rm{d0}}$  for isolated rigid spherical particle can be calculated by the Schiller and Nauman (1935) equation
\begin{equation}
C_{\rm{d0}} =\left\{
\begin{aligned}
&\frac{{24}}{{R{e_{\rm{p}}}}}\left( {1 + 0.15Re_{\rm{p}}^{0.687}} \right){\rm{,}} & R{e_{\rm{p}}} < 1000,\\
&0.44 {\rm{,}} & R{e_{\rm{p}}} \ge 1000.
\end{aligned}\right.
\end{equation}
Here the particle Reynolds number is defined as follows:
\begin{equation}
  R{e_{\rm{p}}} = \frac{{{\varepsilon _{\rm{g}}}{\rho _{\rm{g}}}\left| {{\bf{u}} - {\bf{v}}} \right|{d_{\rm{p}}}}}{{{\mu _{\rm{g}}}}}
\end{equation}

It has been demonstrated that the drag correction factor $\omega$, varying in value between 0.0152 and 4.2876, plays an important role in simulation of gas-solid fluidization (Yang et al., 2003), and hence is employed in a similar way to reflect the heterogeneous structures. The modified LBE is thus written as
\begin{equation}
 \begin{split}
{f_i}({\bf{x}} + {{\bf{c}}_i}\Delta t,&t + \Delta t) = {f_i}({\bf{x}},t)\\
&- \frac{1}{{{\tau _{\rm{t}}}}}\Big(1 - \gamma ({\varepsilon _{\rm{s}}},{\tau _{\rm{t}}})\Big)\Big({f_i}({\bf{x}},t) - {f_i}^{{\rm{eq}}}({\bf{x}},t)\Big)\\
&+ \omega \gamma ({\varepsilon _{\rm{s}}},\tau )\Omega _i^{\rm{s}}
\end{split}
\end{equation}

Accordingly, the volume-averaged gas velocity must be redefined as (Guo and Zhao, 2002; Guo et al., 2002)
\begin{equation}
  \rho {\bf{u}} = \sum\limits_{i = {\rm{0}}}^8 {{{\bf{c}}_i}{\Big(f_i+\frac{1}{2}\omega\gamma\Omega _i^{\rm{s}}}\Big)}
\end{equation}
It should be stressed here that the gas velocity calculated from the modified LBE is the superficial gas velocity and needs to be converted into the actual gas velocity.

\begin{figure*}[htb]
  \centering
  \includegraphics[width=14.3cm]{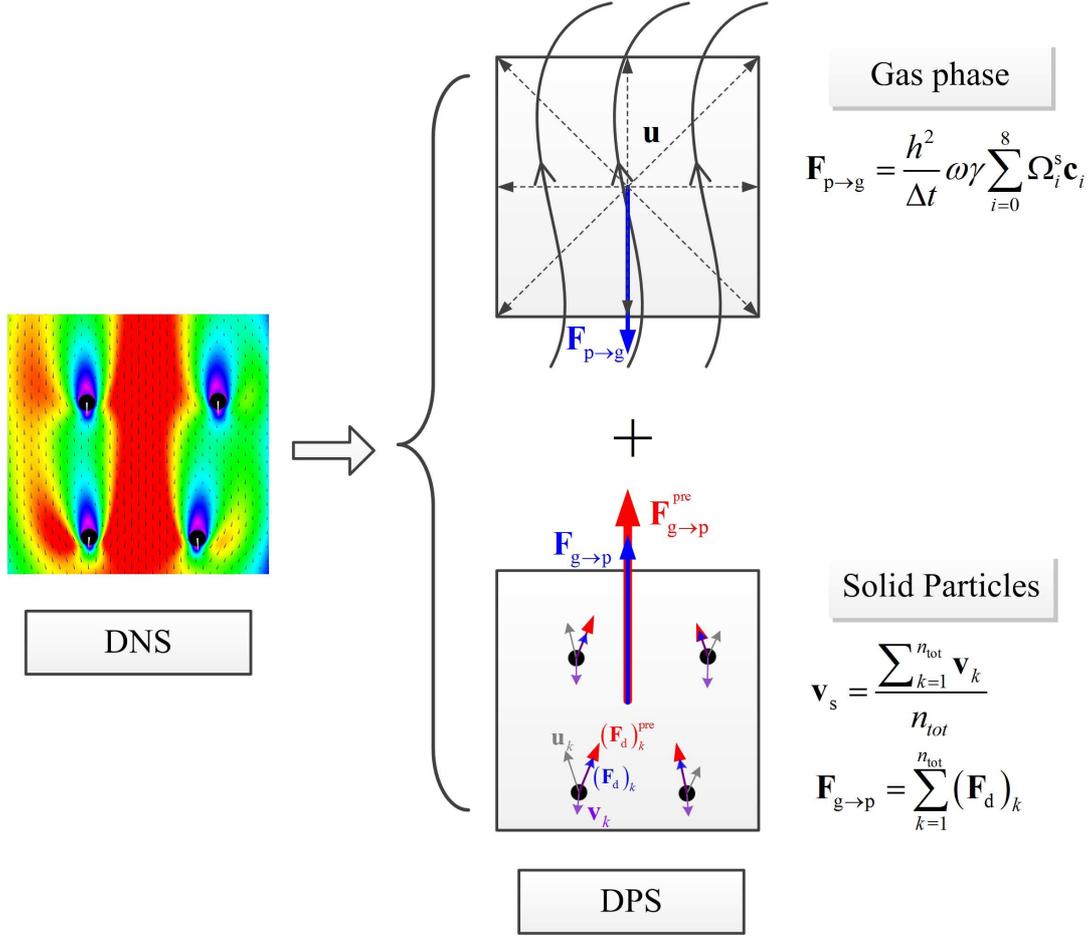}
  \caption{The schematic diagram of gas-solid coupling and drag force correction in a computational cell.}\label{fig3}
\end{figure*}

The hydrodynamic force acting on the fluid in a computational cell, ${{\bf{F}}_{{\rm{p}} \to {\rm{g}}}}$, can be calculated by summing up the momentum transfer that occurs in all discrete directions as
\begin{equation}
  {{\bf{F}}_{{\rm{p}} \to {\rm{g}}}} = \frac{{{h^2}}}{{\Delta t}}\omega \gamma \sum\limits_{i = {\rm{0}}}^8 {\Omega _i^{\rm{s}}{{\bf{c}}_i}}
\end{equation}
Then a pre-estimated drag force acting on solid particle\textit{k}, ${\left( {{{\bf{F}}_{\rm{d}}}} \right)_k^{\rm{pre}}}$, can be calculated according to the following equation (Hoomans et al., 1996; Li and Kuipers, 2003):
\begin{equation}
  {\left( {{{\bf{F}}_{\rm{d}}}}\right)_k^{\rm{pre}}} = \frac{{{{\left( {{V_{\rm{p}}}} \right)}_k}{\beta _k}}}{{1 - {\varepsilon _k}}}\left( {{{\bf{u}}_k} - {{\bf{v}}_k}} \right)
\end{equation}
Here the local inter-phase momentum transfer coefficient $\beta_k$, the local gas velocity ${{\bf{u}}_k}$ and the local voidage $\varepsilon_k$  are all calculated by bilinear interpolation using the values of four surrounding grid nodes. In each computational cell (see Fig. 3), the pre-estimated total drag force acting on all the particles, ${{\bf{F}}_{{\rm{g}} \to {\rm{p}}}^{\rm{pre}}}$, can be obtained as
\begin{equation}
  {{\bf{F}}_{{\rm{g}} \to {\rm{p}}}^{\rm{pre}}} = \sum\limits_{k{\rm{ = }}1}^{{n_{{\rm{tot}}}}} {{{\left( {{{\bf{F}}_{\rm{d}}}} \right)}_k^{\rm{pre}}}}
\end{equation}
which is not necessarily equal to the hydrodynamic force acting on the fluid. To satisfy Newton's third law, the pre-estimated drag force acting on solid particle  \textit{k}, ${\left( {{{\bf{F}}_{\rm{d}}}} \right)_k^{\rm{pre}}}$, is then adjusted to
\begin{equation}
\begin{split}
 &{\left( {{{F}_{\rm{d}}^{\rm{x}}}} \right)_k} = \frac{{\left| {{{F}_{{\rm{p}} \to {\rm{g}}}^{\rm{x}}}} \right|}}{{\left| {{{F}_{{\rm{g}} \to {\rm{p}}}^{\rm{pre,x}}}} \right|}}\frac{{{{\left( {{V_{\rm{p}}}} \right)}_k}{\beta _k}}}{{1 - {\varepsilon _k}}}\left( {{{u}_k^{\rm{x}}} - {{v}_k^{\rm{x}}}} \right),\\
 &{\left( {{{F}_{\rm{d}}^{\rm{y}}}} \right)_k} = \frac{{\left| {{{F}_{{\rm{p}} \to {\rm{g}}}^{\rm{y}}}} \right|}}{{\left| {{{F}_{{\rm{g}} \to {\rm{p}}}^{\rm{pre,y}}}} \right|}}\frac{{{{\left( {{V_{\rm{p}}}} \right)}_k}{\beta _k}}}{{1 - {\varepsilon _k}}}\left( {{{u}_k^{\rm{y}}} - {{v}_k^{\rm{y}}}} \right)
\end{split}
\end{equation}
where $\left( {{{F}_{\rm{d}}^{\rm{x}}}} \right)_k$ and $\left( {{{F}_{\rm{d}}^{\rm{y}}}} \right)_k$  are the x- and y-components of the drag force acting on solid particle $\left( {{{\bf{F}}_{\rm{d}}}} \right)_k$, respectively. The total drag force acting on all the particles, ${{\bf{F}}_{{\rm{g}} \to {\rm{p}}}}$, can be obtained
\begin{equation}
  {{\bf{F}}_{{\rm{g}} \to {\rm{p}}}} = \sum\limits_{k{\rm{ = }}1}^{{n_{{\rm{tot}}}}} {{{\left( {{{\bf{F}}_{\rm{d}}}} \right)}_k}}
\end{equation}
With this drag force correction, the value of the total drag force acting on all the particles automatically equals to the value of the hydrodynamic force acting on the fluid in each computational cell and the Newton's third law, ${{\bf{F}}_{{\rm{g}} \to {\rm{p}}}}={{\bf{F}}_{{\rm{p}} \to {\rm{g}}}}$, is also automatically satisfied in Eqs. (24) and (29).

\subsection{Computational algorithm and implementation}

In traditional DPS, the drag force acting on an individual solid particle is first calculated by the empirical formula Eq. (26), then the total drag force acting on all the particles in each computational cell is calculated by Eq. (27), and the total drag force, as a source term of force, is added into the momentum equation for the fluid phase, which is finally solved to obtain the fluid phase velocity field.

The procedure in this work is slightly different from the case that the fluid phase velocity field is first obtained from the evolution of the modified LBE, the total drag force acting on all the particles is simultaneously calculated by Eq. (25), and then the total drag force is distributed to each particle in the computational cell with the corrections described in Eqs. (26)$\sim$(28). In fact, the pre-estimated drag force acting on the solid particles plays the role of a weight function for this distribution. The step-by-step procedure for the lattice Boltzmann based discrete simulation is outlined in Fig.4, which mainly includes the following steps:

\begin{figure}
  \centering
  \includegraphics[width=8.1cm]{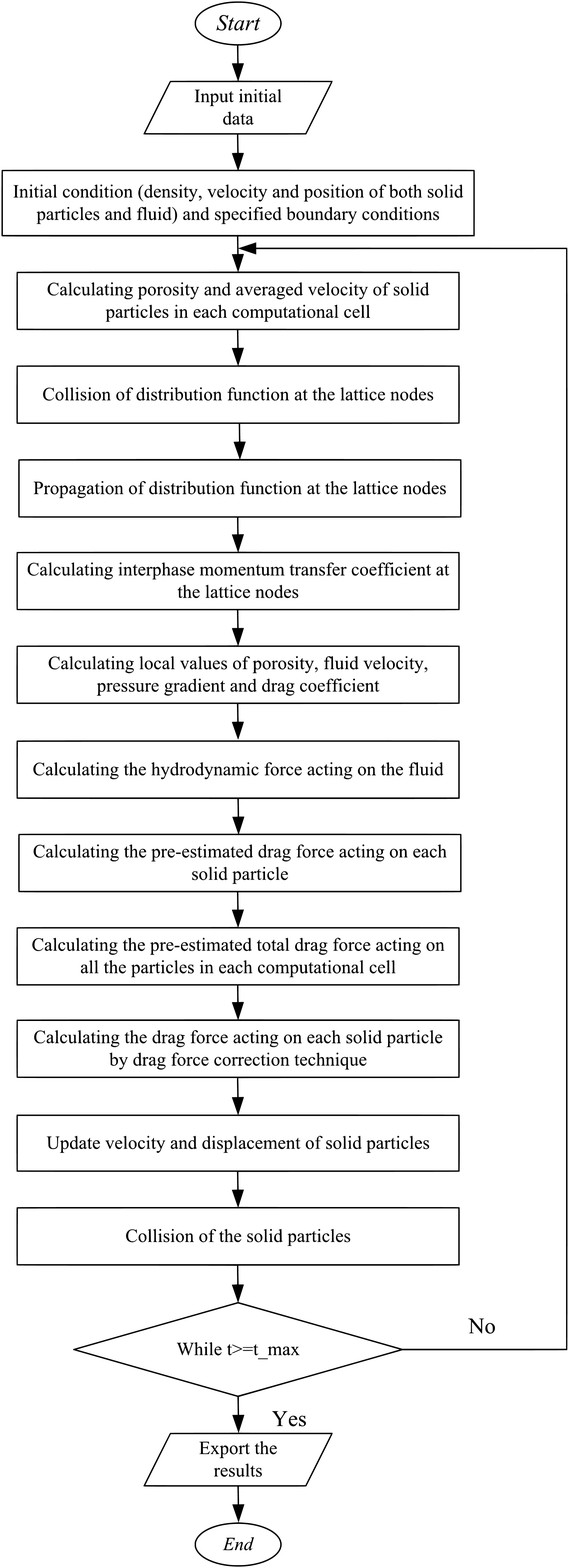}
  \caption{The flow chart of lattice Boltzmann based discrete particle simulation.}\label{fig4}
\end{figure}

\begin{enumerate}[1)]
  \item Input the initial data such as the size and geometry of the simulation domain (for both the solid and fluid phases), the specified boundary conditions (i.e. particle distribution, initial flow field, velocity profile, etc.). The wall boundary condition given in this paper is the bounce-back scheme for fluid and specular reflections for solid particles (Wang et al., 2006).
  \item Perform the statistical calculations of the average velocity and the porosity of solid phase at the fluid lattice nodes according to Eq. (8) and Eq. (11), respectively.
  \item Perform the consecutive propagation and collision process over a discrete fluid lattice cell, and then evolve the fluid flow according to the modified LBE (Eq. (23)), finally obtain the gas flow field and pressure filed.
  \item The local gas velocities, the local pressure gradient, the local porosity, and the local inter-phase momentum transfer coefficient at the center of the particle are calculated by bilinear interpolation using the values of the surrounding computational lattices. Compute ${{\bf{F}}_{{\rm{p}} \to {\rm{g}}}}$  from Eq. (25) and ${{\bf{F}}_{{\rm{g}} \to {\rm{p}}}^{\rm{pre}}}$  from Eqs. (26) and (27) at all lattice nodes. Substitute these values of ${{\bf{F}}_{{\rm{p}} \to {\rm{g}}}}$ and  ${{\bf{F}}_{{\rm{g}} \to {\rm{p}}}^{\rm{pre}}}$ into Eq. (28), obtain the drag force acting on the each particle ${\left( {{{\bf{F}}_{\rm{d}}}} \right)_k}$.
  \item Integrate the particle motion equation (Eq. (16)) numerically to calculate the velocities and the trajectories of individual particles.
  \item Detect the instantaneous collisions of solid particles according to new particle positions and velocities. If the instantaneous collisions occur, the post-collision velocities of individual particles are determined according to Eqs. (17) and (18).
  \item Evolve the next time and go back to step 2 unless the set time step is reached.
\end{enumerate}

\section{Simulations and results }
\label{}
To validate the proposed model, two typical fluidization processes, namely, a single bubble injected into a fluidized bed at incipient fluidization condition and particle clustering in the riser of a circulating fluidization bed, are simulated and compared with published correlations and experimental data. To simplify the simulation, the gas weight is neglected, but the effective gravity acting on the particles is corrected as $(1.0-\rho_{\rm{g}}/\rho_{\rm{s}})\bf{g}$.

\subsection{A single bubble injected into a fluidized bed at incipient fluidization}
\label{}
In this example, a pulse of gas is injected at the bottom of a fluidized bed at incipient fluidization, the gas velocity on both sides of the central orifice is set as the minimum fluidization velocity (background gas velocity) $v_{\rm{mf}}$, and the injection lasts 0.006 seconds. The parameters used for the simulation are summarized in Table 1.

\begin{table*}[htb]
\renewcommand{\arraystretch}{1.2}
\tabcolsep=3.5mm
\caption{Simulation and physical parameters used for bubbling simulation} \label{CFB} \vspace{0.5ex}
\begin{center}
\begin{tabular*}{120mm}{@{\extracolsep{\fill}}*{3}{l}}
\toprule
Item &	Dimensional &	Dimensionless\\
\midrule
Bed width $W(\rm{m})$	& 	0.00675& 	25\\
Bed height $H(\rm{m})$	& 	0.027& 	100\\
Orifice width $r_{\rm{inj}}(\rm{m})$	& 	0.00081& 	3\\
Solid particle diameter $d_{\rm{p}}(\rm{m})$	& 	$5.4\times10^{-5}$& 	0.2\\
Cell or lattice size $h(\rm{m})$	& 	$2.7\times10^{-4}$& 	1\\
Time step $t(\rm{s})$	& 	$1.35\times10^{-6}$& 1	\\
Minimum fluidization velocity $v_{\rm{mf}}(\rm{m/s})$	& 	0.0025& $1.25\times10^{-5}$	\\
Injection velocity $v_{\rm{inj}}(\rm{m/s})$	& 	0.8 & 	0.04\\
Gas density $\rho_{\rm{g}}(\rm{kg/m^{3}})$	& 	1.1795 & 1	\\
Solid density $\rho_{\rm{s}}(\rm{kg/m^{3}})$	& 	930.0 & 788.5	\\
Gas viscosity $\mu_{\rm{g}}(\rm{kg/(m{\kern 1pt}s)})$	& 	$1.8872\times10^{-5}$ & $2.96\times10^{-4}$\\
Coefficient of restitution $e$	& 0.9	& 	0.9\\
Particle number $N$	& 15000	& 	15000\\
\bottomrule
\end{tabular*}
\end{center}
\end{table*}

Figure 5 presents the snapshots of bubble formation at different times. We observe the growth of the spherical cap bubble from an initially small perturbation in the bulk of the bed (see Fig. 5(a)$\sim$(e)) to its subsequent detachment and rising (see Fig. 5(f)$\sim$(h)). Eventually, the bubble wake is followed when the kidney-bubble detaches from the bottom of the bed (see Fig. 5(j)$\sim$(n)). Finally, the bubbles arrive at the surface, and it lifts some particles from wake into the bubble and causes the break-up of the bubble (see Fig. 5(o)$\sim$(p)). This phenomenon is in qualitative agreement with the previous simulation and experimental results of bubbling (Rowe and Yacono, 1976; Nieuwland et al., 1996). It should be noted that the rising bubble shape is more spherical-cap than the elliptic bubble shape in the experimental and numerical results of Bokkers et al. (2004) due to the smaller solid particle diameter.

\begin{figure*}
  \centering
  \includegraphics[width=13cm]{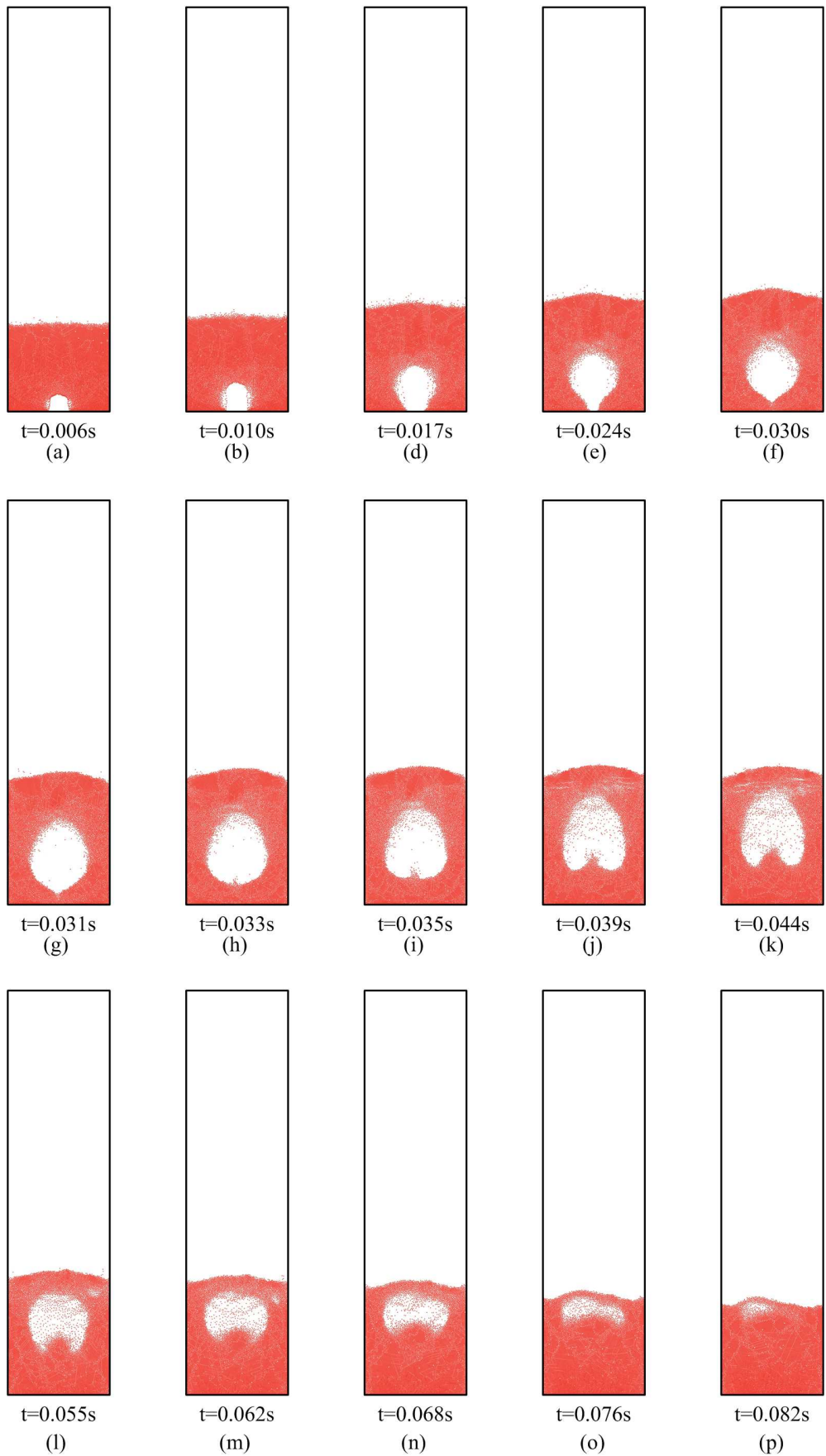}
  \caption{Evolution of bubble formation at a single orifice.}\label{fig5}
\end{figure*}

Figure 6 shows that the gas phase leaves the roof of bubble, and the downward moving particles near the wall drag the gas to the bottom of the bubble where it re-enters the bubble region, resulting in a pair of symmetrical vortices which are observed in the neighborhood of the rising bubble. In addition, the simulated fluidizing gas streamlines passing through a rising bubble are close to those predicted by the two-phase model of a fluidized bed (Davidson and Harrison, 1963). These results illustrate that the modified LBE is capable of capturing accurately the detailed flow structure of the gas phase.

The volume-averaged equivalent bubble diameter is defined as the diameter of a circle with equal area to the region where the gas voidage is higher than 0.85 (Nieuwland et al., 1996). The computed maximum bubble diameter $D_{\rm{e}}$ at different injection velocities is shown in Fig. 7, together with the correlations of Cai et al. (1994). The results show that $D_{\rm{e}}$ increases with the increasing of injection velocities (or equivalently overall superficial gas velocity), and the deviation is below 5\%, suggesting that the simulation results are reasonable.
\begin{figure}[htb]
  \centering
  \includegraphics[width=8.3cm]{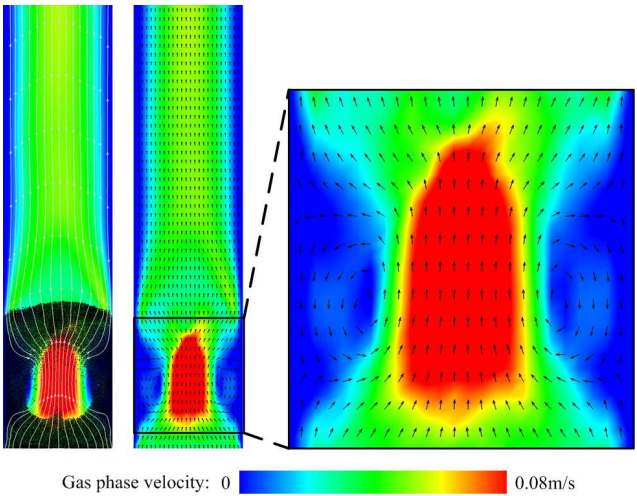}
  \caption{Snapshots of the flow field of gas phase at simulation time \textit{t}=0.035s.}\label{fig6}
\end{figure}

\begin{figure}[htb]
  \centering
  \includegraphics[width=8.3cm]{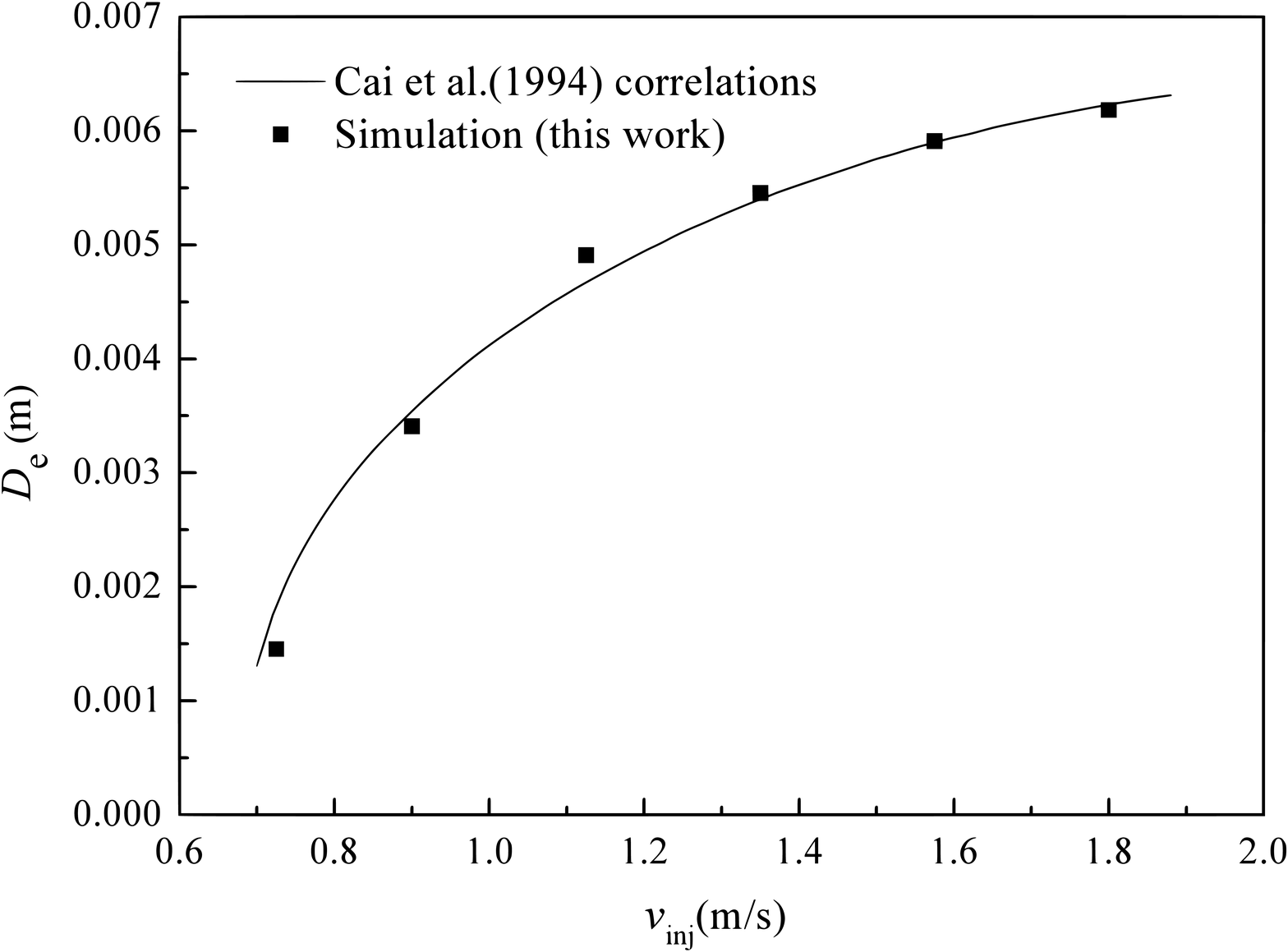}
  \caption{Comparison of the correlation formula (Cai et al., 1994) obtained and simulated maximum bubble diameter for a range of injection velocities.}\label{fig7}
\end{figure}

\subsection{Riser behavior of fluidization systems}
\label{}
The riser hydrodynamics of circulating fluidized beds (CFB) have been investigated extensively in both experiments and simulations in past decades. Our simulation refers to a CFB established by Li and Kwauk (1994), for which a large number of experimental data have been collected. The 2D riser configuration has been reported enough to mimic the real system (Cammarata et al., 2003; Xie et al., 2008), and the comparison of planar 2D simulation results with cylindrical 3D experimental data could also be found widely in the literature to study gas-solid fluidized beds (Wang et al., 2008; Lan et al., 2009; Lu et al., 2011; Chen et al., 2012; Li et al., 2012). Table 2 lists the parameters used in this present simulation.

\begin{table*}
\renewcommand{\arraystretch}{1.2}
\tabcolsep=3.5mm
\caption{Simulation and physical parameters used for riser of circulating fluidized beds} \label{riser} \vspace{0.5ex}
\begin{center}
\begin{tabular*}{120mm}{@{\extracolsep{\fill}}*{3}{l}}
\toprule
Item &	Dimensional &	Dimensionless\\
\midrule
Bed width $W(\rm{m})$	& 	0.0162& 	60\\
Bed height $H(\rm{m})$	& 	0.194& 	720\\
Initial solids volume fraction $\theta_{\rm{s}}$	& 	0.09007 & 0.09007 \\
Solid particle diameter $d_{\rm{p}}(\rm{m})$	& 	$5.4\times10^{-5}$& 	0.2\\
Cell or lattice size $h(\rm{m})$	& 	$2.7\times10^{-4}$& 	1\\
Time step $t(\rm{s})$	& 	$6.0\times10^{-6}$& 1	\\
Gas velocity at the inlet $v_{\rm{inlet}}(\rm{m/s})$ & 1.52	 & 	0.0338\\
Gas density $\rho_{\rm{g}}(\rm{kg/m^{3}})$	& 	1.1795 & 1	\\
Solid density $\rho_{\rm{s}}(\rm{kg/m^{3}})$	& 	930.0 & 788.5	\\
Gas viscosity $\mu_{\rm{g}}(\rm{kg/(m{\kern 1pt}s)})$	& 	$1.8872\times10^{-5}$ & $1.32\times10^{-4}$\\
Coefficient of restitution $e$	& 0.9	& 	0.9\\
Particle number $N$	& 123921	& 	123921\\
\bottomrule
\end{tabular*}
\end{center}
\end{table*}

\begin{figure*}
  \centering
  \includegraphics[width=14.3cm]{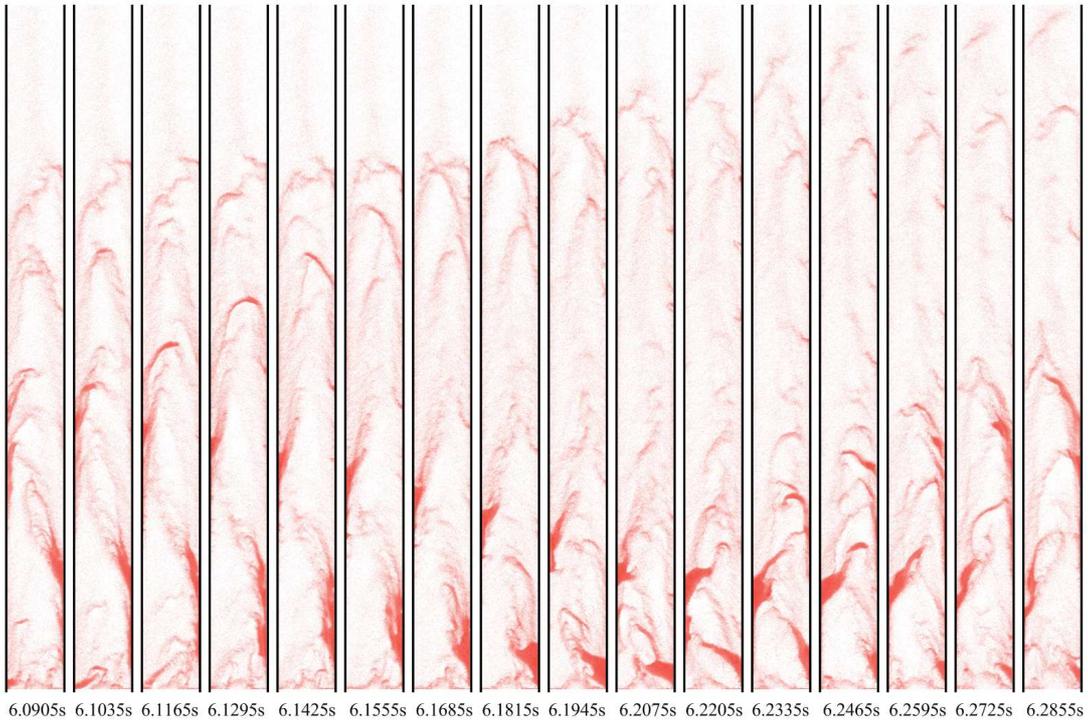}
  \caption{Evolution of flow structures in the riser of circulating fluidized beds.}\label{fig8}
\end{figure*}

Figure 8 shows some snapshots from this simulation. Apparently, radial and axial heterogeneous structures are formed spontaneously and gradually from the homogeneous initial state and appear all the time hereafter.

Figure 9 shows the evolution of the simulated solids flux with time and its comparison with experimental data of Li and Kwauk (1994). Obviously, the simulated solids fluxes $\rm{(13.8kg/(m^{2}s))}$ are in good agreement with the experimental value $\rm{(14.3kg/(m^{2}s))}$.

\begin{figure}
  \centering
  \includegraphics[width=8.3cm]{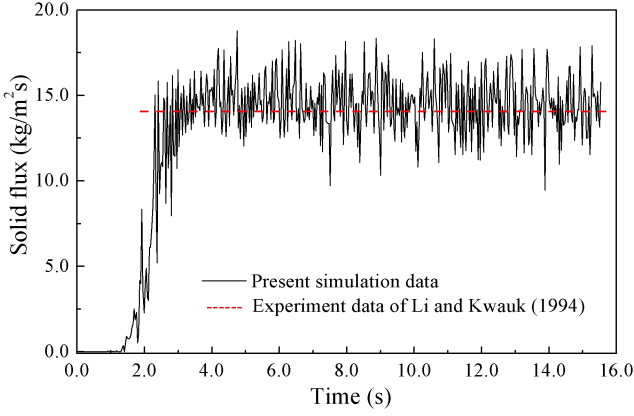}
  \caption{Evolution of the outlet solid flux in the riser of circulating fluidized beds.}\label{fig9}
\end{figure}

Figure 10 shows the comparison of the axial voidage profiles between the simulated results and the experimental data of Li and Kwauk (1994). The simulated profiles are calculated based on time-averaged from 4.0s$\sim$6.6s. The predicted sigmoid distribution of voidage in the axial direction is in reasonable agreement with experimental results, and the deviation may be partly ascribed to the unrealistic setup of inlet and outlet boundary conditions in the 2D simulation.
\begin{figure}
  \centering
  \includegraphics[width=3.3cm]{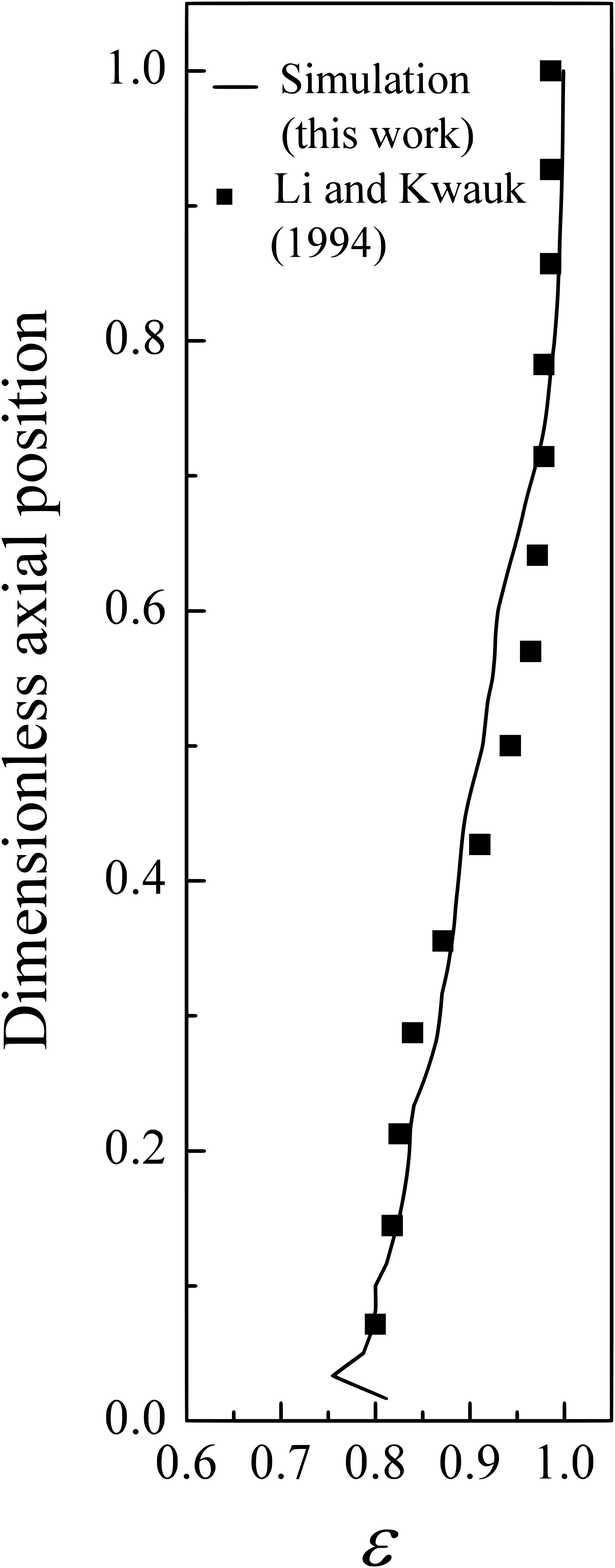}
  \caption{Axial voidage profile in the riser of circulating fluidized beds.}\label{fig10}
\end{figure}

\begin{figure}
  \centering
  \includegraphics[width=8.3cm]{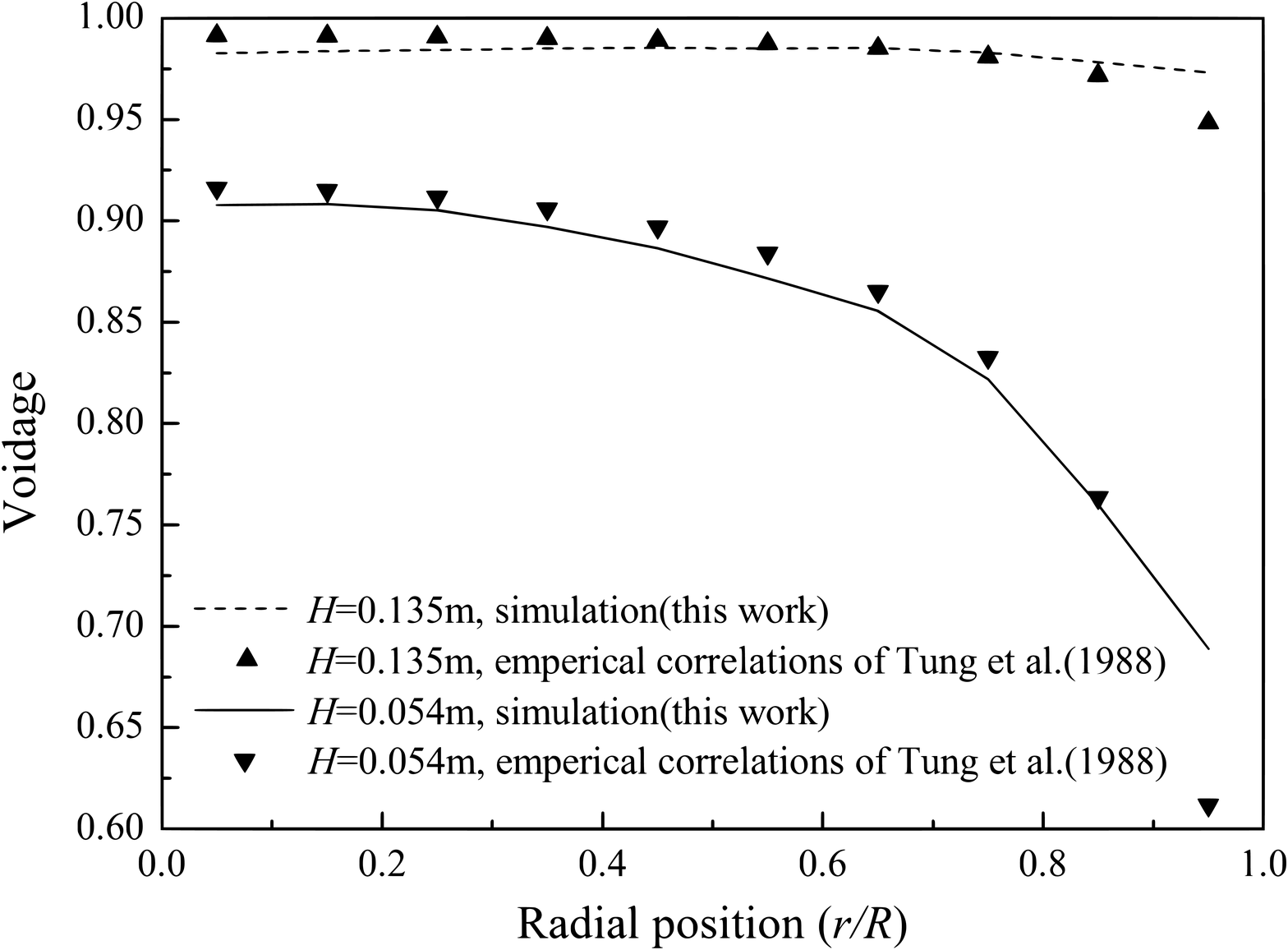}
  \caption{ Radial voidage profiles in the riser of circulating fluidized beds.}\label{fig11}
\end{figure}

Figure 11 shows the comparison of the radial voidage profiles between the simulated results and the experimental correlations proposed by Tung et al. (1988). The simulated profiles are also calculated based on time-averaged from 4.0s$\sim$6.6s, which are in good agreement with empirical correlations in the core region, but overpredict the voidage in the annulus region. The coexistence of a dense annulus and a dilute core can be recognized, and the radial profile in the top region is larger than that in the bottom region. This implies that we have successfully captured a region with the coexistence of a dilute section at the top and a dense section at the bottom.

\section{Discussion and conclusions}
\label{}
In this paper, a new numerical modeling method is presented for discrete particle simulation of gas-solid fluidization; while the lattice Boltzmann method is applied to model the gas flow at a scale larger than the particles, the time-driven hard-sphere model is employed to describe the motion of individual particles and the EMMS drag is used to correct gas-solid interaction. The bubble formation of a single jet into the fluidized bed and the heterogeneous flow structures in the riser of a circulating fluidized bed are simulated successfully by the proposed approach.

Although the simulated results of the proposed approach seem reasonable, the current investigation is still very preliminary. The following aspects need to be clarified further:
\begin{itemize}
  \item \emph{The corresponding macroscopic equations of the modified LBE.} At present, the immersed boundary method is introduced to the framework of LBM by an additional collision term to ensure that the LBE varies smoothly between the nodes occupied by pure fluid or by pure solid. This is a convenient approximation but lacks rigorous theoretical derivation. The exact form of the LBE presenting the volume-averaged N-S equation for the gas flow in DPS is not yet to be established.
  \item \emph{Gas-solid coupling.} The drag force between gas-solid may be still a source of controversy in DPS due to the uncertainty and the variety of different options. Although the present work only considers no-slip discrete force ($\gamma\Omega_i^{\rm{s}}$) with the EMMS drag correction factor ($\omega$) as a source term (Eq. (23)), it is very flexible to incorporate other source terms or corrections into the framework of LBE. In addition to $\omega$, which accounts for the effect of non-uniform particle distribution, the gas velocity distribution below the lattice scale, which is fully resolved in DNS but neglected in this work, may have to be accounted for in a simplified way.
  \item \emph{Fully three-dimensional simulation.} For simplicity of implementation, the current simulation lies in the restriction of two-dimensional fluid and three-dimensional individual spherical particles, and then it can be regarded as a quasi-three-dimensional simulation. Further, its extension to three-dimensions is straightforward in principle, namely using the three-dimensional lattice Boltzmann models (i.e. D3Q15, D3Q19 and D3Q27) (d'Humieres et al., 2002) for gas flows. For three-dimensional physical problems of interest, despite the relatively high numerical efficiency of LBM and TDHS, excessive computational cost
      is still required. Naturally, we will resort to parallel computation due to the advantage of the natural parallelisms inherent in LBM and TDHS. Therefore, implementing three-dimensional massive parallel computation of the proposed model is highly desirable in future.
  \item \emph{Turbulence.} Two-dimensional LES just provides a starting point for modeling high Re number gas flows and probably makes no sense here since the turbulence is inherently three-dimensional in fluidized beds.
\end{itemize}

In closing, LBM preserves the advantages of explicit methods in terms of fast speed and parallelism. While, time-driven hard-sphere model, which can be regarded as a combination of molecular dynamics (MD) and direct simulation Monte Carlo (DSMC) (Bird, 1994), inherits the virtue of computational efficiency in DSMC and physical picture in MD. Therefore, it can be expected that the proposed approach, combining LBM and TDHS, should be a fast numerical method, especially for highly parallel computing. In addition, we would like to point out that the proposed LBM based fluid flow solver can be combined with other solid phase solvers, such as DEM, DSMC (Liu and Lu, 2009), smoothed particle hydrodynamics (SPH) (Gingold and Monaghan, 1977; Monaghan, 1992; Xiong et al., 2011), and particle-in-cell (PIC) (Snider, 2001). We would like to extend the present method to complex cases involving non-spherical particles in the future work.

\section*{Notation}
\begin{deflist}[A]\small
\defitem{$c$}\defterm{lattice speed, m/s}
\defitem{$\textbf{c}$}\defterm{unit velocity of lattice, m/s}%
\defitem{$C_{\rm{d0}}$}\defterm{drag coefficient, kg/($\rm{m}^3${\kern 1pt}s)}
\defitem{$c_{\rm{s}}$}\defterm{speed of sound, m/s}
\defitem{$C_{\rm{s}}$}\defterm{Smagorinsky constant number}
\defitem{$d_{\rm{p}}$}\defterm{solid particle diameter, m}
\defitem{$e$}\defterm{coefficient of restitution}
\defitem{$f$}\defterm{distribution functions}
\defitem{$\mathbf{F}$}\defterm{force, m/$\rm{s}^2$}
\defitem{$f_i^{\rm{eq}}$}\defterm{particle equilibrium distribution function}
\defitem{$\mathbf{g}$}\defterm{gravitational acceleration, m/$\rm{s}^2$}
\defitem{$h$}\defterm{cell or lattice size, m}
\defitem{$H$}\defterm{bed height, m}
\defitem{$m$}\defterm{mass of solid particle, kg}
\defitem{$n_{\rm{tot}}$}\defterm{number of particles in a cell}
\defitem{$N$}\defterm{total number of solid particles in the simulation}
\defitem{$\mathbf{P}$}\defterm{position of solid particle, m}
\defitem{$Q_{ij}$}\defterm{momentum fluxes}
\defitem{$r$}\defterm{radius of solid particle, m}
\defitem{$r_{\rm{inj}}$}\defterm{orifice width, m}
\defitem{$Re_{\rm{p}}$}\defterm{particle Reynolds number}
\defitem{$S_{ij}$}\defterm{strain rate tensor, $\rm{s}^{-1}$}
\defitem{$t$}\defterm{time, s}
\defitem{$\Delta t$}\defterm{time step, s}
\defitem{$\mathbf{u}$}\defterm{gas velocity, m/s}
\defitem{$\mathbf{v}$}\defterm{particle velocity, m/s}
\defitem{$v_{\rm{inj}}$}\defterm{injection velocity, m/s}
\defitem{$v_{\rm{inlet}}$}\defterm{gas velocity at the inlet, m/s}
\defitem{$v_{\rm{mf}}$}\defterm{minimum fluidization velocity, m/s}
\defitem{$V_{\rm{p}}$}\defterm{volume of particle, $\rm{m}^3$}
\defitem{$W$}\defterm{bed width, m}
\defitem{$\mathbf{x}$}\defterm{position of lattice, m}
\defitem{$\Delta x$}\defterm{grid size, m}

\subsubsection*{Greek letters}
\vspace{3.5mm}
\defitem{$\beta$}\defterm{drag coefficient, kg/($\rm{m}^3${\kern 1pt}s)}
\defitem{$\varepsilon$}\defterm{voidage}
\defitem{$\varepsilon_{\rm{g}}$}\defterm{gas volume fractions}
\defitem{$\varepsilon_{\rm{s}}$}\defterm{solid volume fractions}
\defitem{$\varepsilon_{\rm{2d}}$}\defterm{two dimensional porosity}
\defitem{$\varepsilon_{\rm{3d}}$}\defterm{three dimensional porosity}
\defitem{$\gamma$}\defterm{weighting function}
\defitem{$\mu_{\rm{g}}$}\defterm{gas viscosity, kg/(m{\kern 1pt}s)}
\defitem{$\nu$}\defterm{kinematic viscosity, $\rm{m}^2$/s}
\defitem{$\theta_{\rm{s}}$}\defterm{initial solids volume fraction}
\defitem{$\rho_{\rm{g}}$}\defterm{gas density, kg/$\rm{m}^3$}
\defitem{$\rho_{\rm{s}}$}\defterm{solid density, kg/$\rm{m}^3$}
\defitem{$\omega$}\defterm{lattice weight or drag correction factor}
\defitem{$\Omega$}\defterm{collision term}
\subsubsection*{Subscripts}
\vspace{3.5mm}
\defitem{$\rm{1,2}$}\defterm{the numbers of solid particles}
\defitem{$\rm{c}$}\defterm{collion}
\defitem{$\rm{d}$}\defterm{drag}
\defitem{$\rm{e}$}\defterm{turbulent eddy or average equivalent}
\defitem{$\rm{g}$}\defterm{gas phase}
\defitem{$i$}\defterm{the \textit{i}th direction of lattice discrete velocity}
\defitem{$k$}\defterm{the \textit{k}th solid particle}
\defitem{$\rm{p}$}\defterm{particle}
\defitem{$\rm{pre}$}\defterm{pre-estimated}
\defitem{$\rm{s}$}\defterm{solid phase}
\defitem{$\rm{t}$}\defterm{total}
\defitem{$\rm{x}$}\defterm{x-direction}
\defitem{$\rm{y}$}\defterm{y-direction}
\subsubsection*{Abbreviations}
\vspace{3.5mm}
\defitem{BGK}\defterm{\quad \quad Bhatnagar-Gross-Krook}
\defitem{CFB}\defterm{\quad \quad Circulating Fluidized Beds}
\defitem{CFBC}\defterm{\quad \quad Circulating Fluidized Bed Combustion}
\defitem{CFD}\defterm{\quad \quad Computational Fluid Dynamics}
\defitem{DEM}\defterm{\quad \quad Discrete Element Method}
\defitem{DPS}\defterm{\quad \quad Discrete Particle Simulation}
\defitem{DNS}\defterm{\quad \quad Direct Numerical Simulation}
\defitem{DSMC}\defterm{\quad \quad Direct Simulation Monte Carlo}
\defitem{D2Q9}\defterm{\quad \quad 2-Dimensional 9-Velocity Lattice Boltzmann Model}
\defitem{D3Q15}\defterm{\quad \quad 3-Dimensional 15-Velocity Lattice Boltzmann Model}
\defitem{D3Q19}\defterm{\quad \quad 3-Dimensional 19-Velocity Lattice Boltzmann Model}
\defitem{D3Q27}\defterm{\quad \quad 3-Dimensional 27-Velocity Lattice Boltzmann Model}
\defitem{EMMS}\defterm{\quad \quad Energy Minimization Multi-Scale}
\defitem{FCC}\defterm{\quad \quad Fluid Catalytic Cracking}
\defitem{LGA}\defterm{\quad \quad Lattice Gas Automata}
\defitem{LBE}\defterm{\quad \quad Lattice Boltzmann Equation}
\defitem{LBM}\defterm{\quad \quad Lattice Boltzmann Method}
\defitem{LES}\defterm{\quad \quad Large Eddy Simulation}
\defitem{MD}\defterm{\quad \quad Molecular Dynamics}
\defitem{PBE}\defterm{\quad \quad Population Balance Equation}
\defitem{PIC}\defterm{\quad \quad Particle-In-Cell}
\defitem{QBMM}\defterm{\quad \quad Quadrature-Based Moment Methods}
\defitem{SGS}\defterm{\quad \quad Sub-Grid Scale}
\defitem{SPH}\defterm{\quad \quad Smoothed Particle Hydrodynamics}
\defitem{TDHS}\defterm{\quad \quad Time-Driven Hard-Sphere}
\defitem{TFM}\defterm{\quad \quad Two-Fluid Model}
\end{deflist}

\section*{Acknowledgement}
\label{}
This work is financially supported by the National Natural Science Foundation of China under Grants No. 21106155, the Specialized Fund from the Youth Innovation Promotion Association of the Chinese Academy of Science and the Chinese Academy of Sciences under Grant No. XDA07080303. We would like to thank Prof. Junwu Wang for his valuable suggestions and Dr. Qingang Xiong for useful discussions on this work.

\bibliographystyle{elsarticle-harv}
\bibliography{<your-bib-database>}



\end{document}